\documentclass[aps,prb,twocolumn,amsmath,amssymb,superscriptaddress]{revtex4-1}

\usepackage{graphicx}
\usepackage{bm}
\usepackage{color}

\usepackage{booktabs}
\AtBeginDocument{
\heavyrulewidth=.08em
\lightrulewidth=.05em
\cmidrulewidth=.03em
\belowrulesep=.65ex
\belowbottomsep=0pt
\aboverulesep=.4ex
\abovetopsep=0pt
\cmidrulesep=\doublerulesep
\cmidrulekern=.5em
\defaultaddspace=.5em
}

\newcommand{\summ}{\displaystyle\sum}
\newcommand{\angstrom}{\mbox{\normalfont\AA}}

\begin{document}


\title{Critical Scaling and Aging near the Flux Line Depinning Transition}

\author{Harshwardhan Chaturvedi}
	\affiliation{Department of Physics \& Center for Soft Matter and Biological Physics (MC 0435), Virginia Tech, 850 West Campus Drive, Blacksburg, Virginia 24061, USA}
	\affiliation{Kaizen Analytix, 2 Ravinia Drive, Atlanta, Georgia 30346, USA}
\author{Ulrich Dobramysl}
	\affiliation{Wellcome Trust / CRUK Gurdon Institute, University of Cambridge, Tennis Court Rd, Cambridge CB2 1QN, United Kingdom}
\author{Michel Pleimling}
	\affiliation{Department of Physics \& Center for Soft Matter and Biological Physics (MC 0435), Virginia Tech, 850 West Campus Drive, Blacksburg, Virginia 24061, USA}
	\affiliation{Academy of Integrated Science (MC 0563), Virginia Tech, 800 West Campus Drive, Blacksburg, Virginia 24061, USA}
\author{Uwe C. T\"auber}
	\affiliation{Department of Physics \& Center for Soft Matter and Biological Physics (MC 0435), Virginia Tech, 850 West Campus Drive, Blacksburg, Virginia 24061, USA}

\date{\today}

\begin{abstract}
We utilize Langevin molecular dynamics simulations to study dynamical critical behavior of magnetic flux lines near the depinning transition in type-II superconductors subject to randomly distributed attractive point defects. 
We employ a coarse-grained elastic line Hamiltonian for the mutually repulsive vortices and purely relaxational kinetics. In order to infer the stationary-state critical exponents for the continuous non-equilibrium depinning transition at zero temperature $T = 0$ and at the critical driving current density $j_c$, we explore two-parameter scaling laws for the flux lines' gyration radius and mean velocity as functions of the two relevant scaling fields $T$ and $j - j_c$. 
We also investigate critical aging scaling for the two-time height auto-correlation function in the early-time non-equilibrium relaxation regime to independently measure critical exponents. 
We provide numerical exponent values for the distinct universality classes of non-interacting and repulsive vortices.
\end{abstract}

\maketitle

\section{Introduction}
The flow of magnetic flux lines in type-II superconductors in the presence of fixed attractive pinning centers represents a paradigmatic example of coherent structures driven through disordered media. Such systems are of prime interest from a theoretical point of view since they exhibit a rich variety of both thermodynamic phases and non-equilibrium steady states that result from the competing energy scales associated with the intrinsic elastic rigidity, mutual interactions, quenched disorder, thermal fluctuations, and external driving current. From an experimental/technological standpoint they are of paramount importance as well, since they emerge in a highly diverse array of physical scenarios, \textit{e.g.} in directed polymers, magnetic flux vortices, charge density waves (CDWs), magnetic domain walls, moving Wigner crystals, and driven membrane sheets \cite{fily2010}. Indeed, the non-linear dynamics of vortex motion in disordered type-II superconductors has been studied extensively \cite{nattermann1990, ioffe1987, nattermann1987, feigelman1989, blatter1994,pleimling2011,pleimling2015} through numerical simulations but also analytically by means of functional renormalization group techniques \cite{chauve2000, giamarchi2001, brazovskii2004, nattermann2000}. Fisher in 1985 via phenomenological arguments posited that the depinning of sliding CDWs may be regarded as a dynamic critical phenomenon where driving force acts as the control parameter and velocity as the associated order parameter \cite{fisher1985}, an idea that has since been successfully extended to several domains beyond CDWs \cite{nattermann1992, narayan1992, narayan1993, ertas1994, chauve2000, chauve2001, le_doussal2002}. Ample evidence for elastic critical depinning has been found both in experiments \cite{duruoz1995, rimberg1995, kurdak1998, parthasarathy2001, higgins1996, ruyter2008, ammor2010, mohan2009} and in numerical studies \cite{dominguez1994, chen2008, chen2008-1, liu2008, guo2009, lv2009, reichhardt2001, reichhardt2002, reichhardt2003, olive2009}, which all observed clear signatures for a continuous (second-order) dynamical phase transition at a critical value of the external drive.

To mention only a few important recent investigations of the critical depinning of vortices in disordered type-II superconductors, Luo and Hu utilized molecular dynamics simulations to study the dynamical scaling of velocity-force curves for flux lines in a three-dimensional embedded space ($d = 3$), obtaining the critical exponents $\beta$ and $\delta$ in both the weak and strong pinning regimes \cite{luo2007}. Fily \textit{et al.} studied depinning for two-dimensional vortex lattices ($d = 2$), and determined $\beta$ and $\delta$ for the scaling relation that governs the velocity-force behavior near the depinning transition \cite{fily2010}. Di Scala \textit{et al.} computed  critical scaling exponents including the growth exponent $\nu$ for the elastic depinning of vortices in two dimensions \cite{scala2012}. Bag \textit{et al.} recently determined critical scaling exponents from experimental data they obtained for 2H-NbS$_2$ single crystals \cite{bag2018}. The two-dimensional critical depinning dynamics, including non-equilibrium relaxation and aging scaling, of skyrmion topological defects in disordered magnetic films has been investigated by Xiong \textit{et al.} \cite{xiong2019} For a comprehensive up-to-date (until 2016) review article on depinning and non-equilibrium phases in various systems, we refer to Ref.~[\onlinecite{reichhardt2016}].

In this present work, we employ an elastic line model to study critical behavior near the depinning transition for vortices in the presence of weak attractive random quenched disorder (point defects) in a three-dimensional system ($d = 3$) with a two-dimensional displacement vector ($N = 2$) \cite{dobramysl2013,dobramysl2014,assi2015,assi2016,chaturvedi2016,chaturvedi2018}. We perform finite-temperature scaling on both steady-state velocity and radius of gyration data and thereby obtain the stationary critical scaling exponents $\beta$, $\delta$, and $\nu$ that characterize the depinning process as a continuous second-order phase transition at zero temperature, finding $\beta$ to be in good agreement with experimental values. In addition, we probe the non-equilibrium aging dynamics in the system by quenching vortices from the high-drive moving lattice state to the critical depinning regime and studying the ensuing two-time vortex line displacement auto-correlations to compute the aging exponent $b$, dynamic exponent $z$, auto-correlation exponent $\lambda_C$, and roughness exponent $\zeta$ in the system.

\section{Model and simulation description}
We model magnetic flux lines in type-II superconductors as mutually repulsive elastic lines in the extreme London limit \cite{nelson1993, das2003} with the effective Hamiltonian or free energy functional
\begin{equation}
\label{Ham}
\begin{split}
H[\mathbf{r}_i] =  \summ_{i=1}^{N}\int_{0}^{L}dz \Bigg[ \frac{\tilde{\epsilon}_1}{2}\left\vert\frac{d\mathbf{r}_i(z)}{dz}\right\vert^2 + U_D(\mathbf{r}_i(z)) & \\
+ \frac{1}{2}\summ_{j\neq i}^{N}V(|\mathbf{r}_i(z) - \mathbf{r}_j(z)|) - \mathbf{F_d}\cdot\mathbf{r}_i(z)\Bigg] &.
\end{split}
\end{equation}
Here $\mathbf{r}_i(z)$ represents the $xy$ position of the $i$th flux line (one of $N = 16$), at height $z$. Model parameters have been chosen to closely match the material properties of YBCO. The elastic line stiffness or local tilt modulus is given by $\tilde{\epsilon}_1 \approx \Gamma^{-2}\epsilon_0\ln(\lambda_{ab}/\xi_{ab})$ where $\Gamma^{-2} = M_{ab}/M_c = 1/25$ denotes the anisotropy parameter and $\epsilon_0\approx1.92\cdot 10^{-6}\mathrm{erg}/\mathrm{cm}$ is the elastic line energy per unit length. $\lambda_{ab} \approx 1200\angstrom$ is the London penetration depth and $\xi_{ab} \approx 10.5\angstrom$ is the coherence length, in the $ab$ crystallographic plane. The in-plane repulsive interaction between any two flux lines is given by $V(r)=2\epsilon_0K_0(r/\lambda_{ab})$, where $K_0$ denotes the zeroth-order modified Bessel function. It effectively serves as a logarithmic repulsion that is exponentially screened at the scale $\lambda_{ab}$. The pinning sites are modeled as smooth potential wells $U_D(\mathbf{r}, z) = -\summ_{\alpha=1}^{N_D}\frac{b_0}{2}p\left[1-\tanh\left(5\frac{|\mathbf{r}-\mathbf{r}_\alpha|-b_0}{b_0}\right)\right]\delta(z-z_\alpha)$, where $N_D = 1116$ indicates the number of pinning sites, $p = 0.05\epsilon_0$ is the pinning potential strength, $b_0 = 35 \angstrom$ is the width of the potential well, while the vector $\mathbf{r}_\alpha$ and coordinate $z_\alpha$ respectively represent the in-plane and vertical positions of pinning site $\alpha$. The Lorentz force exerted on the flux lines by an external electrical current density $\mathbf{j}$ is modeled in the system as a tunable, spatially uniform drive $F_d = |\mathbf{j} \times \phi_0 \mathbf{B}/B|$ in the $x$ direction where $\phi_0 = hc/2e$ represents the magnetic flux quantum and $\mathbf{B}/B$ is a unit vector pointing in the direction of the magnetic flux. All lengths are expressed in units of $b_0$ while energies are expressed in units of $\epsilon_0 b_0$.

We enforce periodic boundary conditions in the $x$ and $y$ directions and free boundary conditions in the $z$ direction. The system size is $X \times Y \times L = 314b_0 \times 272b_0 \times 100b_0$; the ratio of $X$ to $Y$ is set to $2/\sqrt{3}$ to ensure that the flux lines equilibrate to a periodic hexagonal Abrikosov lattice in the absence of defects.

We simulate the dynamics of the model by discretizing the Hamiltonian (\ref{Ham}) into $L = 100$ layers along the $z$ direction and using it to obtain coupled overdamped Langevin equations 
\begin{equation}
\nonumber
\label{lan}
\eta\frac{\partial\mathbf{r}_i(t, z)}{\partial t}=-\frac{\delta H[\mathbf{r}_i(t, z)]}{\delta\mathbf{r}_i(t, z)}+\mathbf{f}_{i}(t, z)\,,
\end{equation}
which are subsequently solved numerically. Here $\eta=\phi_0^2/2\pi\rho_n c^2 \xi_{ab}^2$ denotes the Bardeen--Stephen viscous drag parameter, where $\rho_n \approx 500\mu\Omega m$ represents the normal-state resistivity of YBCO near $T_c$ \cite{blatter1994, bardeen1965}. This results in the simulation time step being defined by the fundamental temporal unit $t_0=\eta b_0/\epsilon_0\approx 18$ ps. We model the fast, microscopic degrees of freedom of the surrounding medium as uncorrelated Gaussian white noise $\mathbf{f}_{i,z}(t)$ with vanishing mean $\langle\mathbf{f}_{i,z}(t)\rangle=0$. Furthermore, these stochastic forces obey the Einstein relation $\langle\mathbf{f}_{i,z}(t) \cdot \mathbf{f}_{j,z'}(s)\rangle = 4\eta k_BT\delta_{ij}\delta_{zz'}\delta(t-s)$ which ensures that the system relaxes to thermal equilibrium with a canonical probability distribution $P[\mathbf{r}_{i,z}]\propto \textrm{exp}(-H[\mathbf{r}_{i,z}]/k_BT)$ in the absence of any external current. The temperature in the simulations is set to $k_B T / \epsilon_0 b_0 = 0.001$ ($T \approx 5$K) and lower.

\section{Measured quantities}
We directly measure four quantities of interest in our model system:
The mean \emph{radius of gyration} $r_g = \sqrt{\langle(\mathbf{r}_{i}(z) - \langle\mathbf{r}_{i}\rangle_z)^2\rangle}$ is the standard deviation of the lateral positions $\mathbf{r}_{i}(z)$ of the points constituting the $i$th flux line, averaged over all the lines. Hence $r_g$ represents a measure of the overall roughness of the vortex lines in the sample. Here $\langle \ldots \rangle_z$ indicates an average over all layers $z$ of a given flux line, while $\langle \ldots \rangle$ denotes an average over layers $z$, over all vortex lines $i$, and over different realizations of disorder and noise. The mean vortex \emph{velocity} in the direction of the drive ($x$ direction) is given by the $x$-component of the vector $\mathbf{v} = \left\langle d\mathbf{r}_{i}(z)/dt \right\rangle$.
\begin{figure}
  \centering
  \includegraphics[width = .6\linewidth]{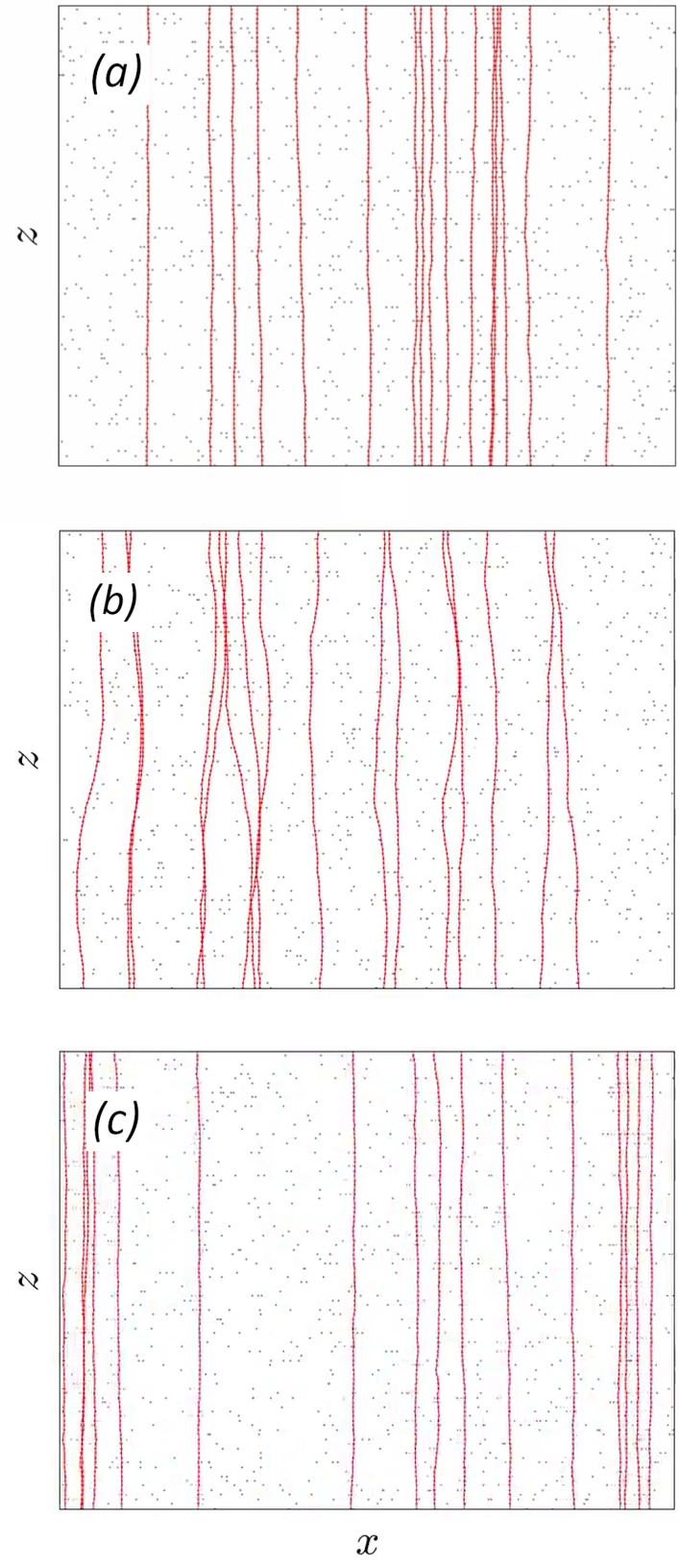}
	\caption{%
	Simulation snapshots of the non-interacting flux line system in the (a) pinned ($F_d=0 \epsilon_0b_0$), (b) critical ($F_d=0.0095 \epsilon_0b_0$), and (c) moving ($F_d=0.03 \epsilon_0b_0$) stationary regimes at temperature $T=0.0009 \epsilon_0b_0/k_B$. The snapshots represent a side view of the system, \emph{i.e.}, a projection of the three-dimensional system onto the $xz$ plane, with drive $F_d$ directed in the positive $x$ direction.
	\label{fig:snapshots}}
\end{figure}
We obtain $r_g$ and $v$ as functions of drive $F_d$ in the \emph{steady state} by randomly placing 16 straight flux lines in the system and immediately subjecting them to thermal fluctuations at temperature $T$ and the desired drive strength $F_d$. The lines are allowed to relax in this constant temperature-drive bath for $100,000 t_0$, until a stationary regime is reached (see Fig.~\ref{fig:snapshots} for snapshots). At this point, we start measuring $r_g$ and $v$ every $100$ time steps, a duration larger than the correlation times in the system. We perform $1,000$ such measurements and record their average for each observable. We simulate $10$ independent realizations and perform an ensemble average. Between the temporal and ensemble averaging, each data point represents a combined mean over $10,000$ independent values.

The third set of quantities measured are normalized two-time vortex ``height'', \textit{i.e.}, transverse flux line displacement auto-correlation functions
\begin{equation}
\nonumber
\label{ha}
C(t,s) = \frac
    { \left< (\mathbf{r}_{i,z}(t) - 
    \left< \mathbf{r}_{i,z}(t) \right>_z)
    (\mathbf{r}_{i,z}(s) - 
    \langle\mathbf{r}_{i,z}(s)\rangle_z)\right>
    }
    {\left< ( \mathbf{r}_{i,z}(s) - 
    \left<\mathbf{r}_{i,z}(s)\right>_z )^2 \right>
    }\,
\end{equation}
that quantify how correlated the lateral positions $\mathbf{r}_{i,z}$ of the elements of a line relative to the mean lateral line position $\langle\mathbf{r}_{i,z}\rangle_z$ at the present time $t$ are to their values at a past time $s$; they measure the time evolution of local transverse thermal vortex fluctuations. We use height auto-correlations to investigate the existence and nature of physical aging in our system. A system shows aging when a dynamical two-time quantity displays slow relaxation and the breaking of time translation invariance \cite{henkel2010}. Additionally, in a \emph{simple aging} scenario, the two-time quantity satisfies dynamical scaling and obeys the general scaling form 
\begin{equation}
\label{eq:aging_ansatz}
C(t,s)=s^{-b}f_C(t/s)\,,
\end{equation}
where $f_C$ is a scaling function that follows the asymptotic power law $f_C(t/s)\sim(t/s)^{-\lambda_C/z}$ as $t\rightarrow\infty$; $b$ is called the aging scaling exponent, $\lambda_C$ the auto-correlation exponent, and $z$ is the dynamical scaling exponent. 

In this study, we measure height auto-correlations following \emph{drive quenches}. A drive quench is performed by first taking the system to a steady state (as described above) at some initial drive strength $F_d$ followed by an instantaneous change (quench) of the drive strength to the desired final value. Following the quench, we wait for some waiting time $s$ before taking a snapshot of the system and proceeding to measure $C(t, s)$ with respect to the snapshot at times $t > s$; this is repeated for several waiting times. All results are averaged over at least $10,000$ realizations of disorder and noise.

Finally, we extract the characteristic system \emph{correlation time} $\tau$ by measuring the time taken for $C(t, s)$ (for arbitrary $s$) to fall from its value $1$ at $t = s$ to $0.5$ at later time $t$.

\section{Stationary critical scaling}
As vortex depinning from attractive point defects represents a zero-temperature non-equilibrium continuous phase transition \cite{fisher1985}, the critical scaling of the $v$--$F_d$ curves above but near the depinning threshold $F_d = F_c$ should be described by a power law $v(T=0, f>0) \sim f^\beta$ where $f = (F_d - F_c)/F_c$ is the reduced force. More generally, the critical behavior in the $(T,f)$ control parameter plane is captured by the scaling ansatz
\begin{equation} \label{eq:v_fin_tem_ans}
v(T, f) = T^{1/\delta} S(T^{-1 / \beta \delta} f)\ ,
\end{equation}
where $S(x)$ is a scaling function that satisfies the conditions $S(x \to  \infty) \sim x^\beta$ and $S(x = 0) = \textrm{const}$ \cite{fisher1983, fisher1985, middleton1992, roters1999, luo2007, fily2010}. Taking the limit $T \to 0^+$ in (\ref{eq:v_fin_tem_ans}) yields the prescribed power law for $v$ as function of $f$ at zero temperature, while setting $f = 0$ yields the algebraic temperature dependence $v(T > 0, f = 0) \sim T^{1/\delta}$.
\begin{figure}
  \centering
  \includegraphics[width = .7\linewidth]{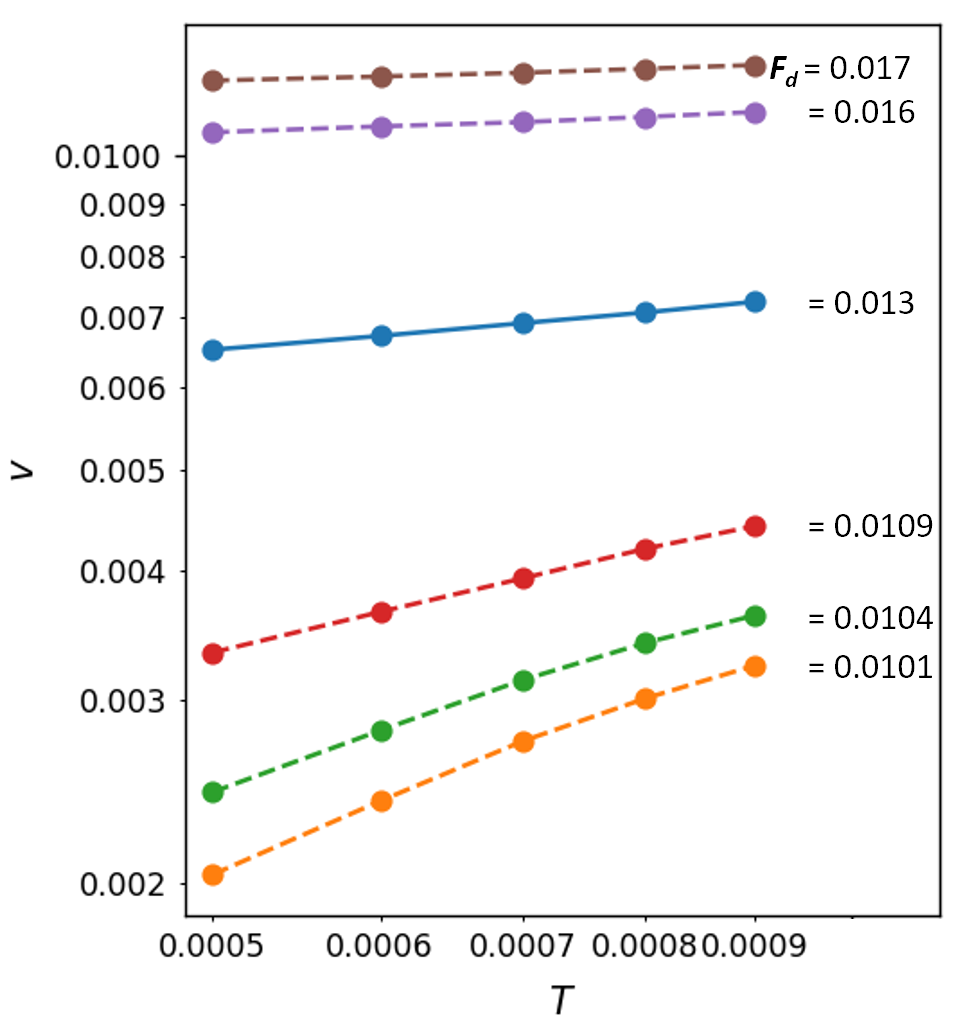}
	\caption{%
	Velocity-temperature ($v$--$T$) curves for repulsive vortex lines taken for various values of the drive $F_d$, with the curve at critical drive $F_c = (0.013 \pm 0.0005) \epsilon_0$ indicated by a solid line.
	\label{fig:v_vs_T}}
\end{figure}
\begin{figure}
  \centering
  \includegraphics[width = .99\linewidth]{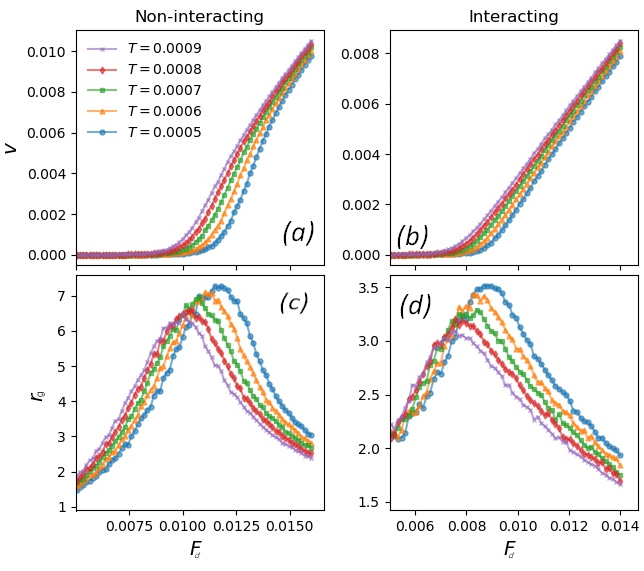}
	\caption{%
	Steady-state (a, b) velocity $v$ (in units of $b_0/t_0$) and (c, d) radius of gyration $r_g$ (in units of $b_0$) as functions of driving force $F_d$ (in units of $\epsilon_0$) for (a, c) non-interacting and (b, d) interacting flux lines. Each quantity is measured at five different temperatures $T$ (values listed in units of $\epsilon_0 b_0/k_B$).
	\label{fig:crit_steadys}}
\end{figure}

We argue that the radius of gyration $r_g$ plays the role of the critical correlation length $\xi$ in the system, and hence upon approaching the transition $f \to 0^+$ should diverge according to $r_g(T=0, f>0) \sim f^{-\nu}$ [\onlinecite{ma2000}]. As with the scaling of the $v$--$f$ curves, we postulate the analogous two-parameter scaling ansatz
\begin{equation} \label{eq:r_fin_tem_ans}
r_g(T, f) = T^{- \nu / \beta \delta} R(T^{-1 / \beta \delta} f) \ ,
\end{equation}
with $R(x \to  \infty) \sim x^{-\nu}$ and $R(x = 0) = \textrm{const}$. Taking $T \to 0^+$ yields the required $v$--$f$ power-law and setting $f = 0$ yields the scaling relation $r_g(T>0, f = 0) \sim T^{-\nu / \beta \delta}$.
Finally, the correlation time $\tau$ is expected to diverge as $\tau \sim f^{-\nu z}$ at $T=0$ near the critical point $f  \to 0^+$ [\onlinecite{ma2000}].
\begin{figure}
  \centering
  \includegraphics[width = .99\linewidth]{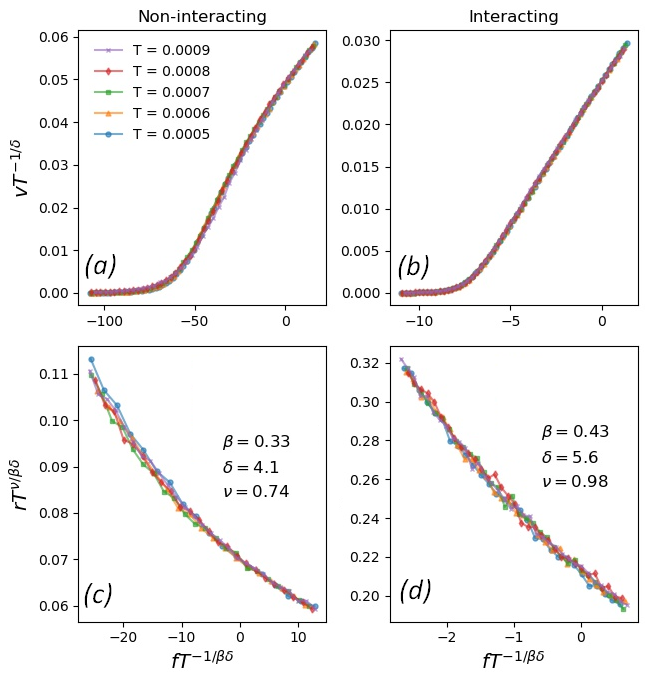}
	\caption{%
	Steady-state data from Fig. \ref{fig:crit_steadys} re-plotted by means of the two-parameter scaling ans\"atze (\ref{eq:v_fin_tem_ans}) and  (\ref{eq:r_fin_tem_ans}) with stationary critical exponents $\beta$, $\delta$, and $\nu$. Panels (a, c) show results for non-interacting vortices,
	while (b, d) represent the data for repulsively interacting flux lines.
	Panels (a, b) show scaled velocities $v$ for five different temperatures $T$ (in units of $\epsilon_0 b_0 / k_B$) as functions of scaled reduced drive $f$, and panels (c, d) display scaled gyration radii $r_g$ for the same temperatures, also as functions of $f$.
	\label{fig:crit_steadys_scaled}}
\end{figure}

To numerically determine the reduced drive $f$, we first find the zero-temperature critical drive $F_c$ via Eq.~(\ref{eq:v_fin_tem_ans}): $v$ should exhibit power law behavior as a function of $T$ when $f=0$ ($F_d=F_c$). Therefore, on a double-logarithmic plot, the $v$--$T$ curves for $F_d<F_c$ are concave, those for $F_d>F_c$ in contrast are convex, and at the critical drive $F_c=(0.013\pm 0.0005) \epsilon_0$ the curves are approximately linear for the interacting system (shown in Fig.~\ref{fig:v_vs_T}). From the slope and via Eq.~(\ref{eq:v_fin_tem_ans}), we find $\delta = 5.6 \pm 0.2$. Identical inflection analysis of $v$--$T$ curves for non-interacting flux lines yields $F_c = (0.015 \pm 0.0005) \epsilon_0$ and $\delta = 4.1 \pm 0.1$.

We show the steady-state velocity and radius of gyration as a function of drive in Fig.~\ref{fig:crit_steadys}. Note that $F_c$ is lower for the interacting system than the non-interacting one; this is consistent with the enabling role played by inter-vortex repulsions in the depinning process that facilitates collective unbinding of correlated flux line clusters.
\begin{table*}[]
\centering
\begin{tabular}{@{}llllll@{}}

Source & $d$ & $\beta\delta$ & $\beta$ & $\delta$ & $\nu$ \\ \midrule[1.2pt]

Luo and Hu \cite{luo2007} (simulation) & $3$ \hspace{1em} & $1.0 \pm 0.019$ \hspace{1em} & $0.754 \pm 0.010$ \hspace{1em} & $1.326 \pm 0.018$ \hspace{1em} &  \\ \midrule

Fily \emph{et al.} \cite{fily2010} (simulation) & $2$ & $1.73 \pm 0.27$ & $1.30 \pm 0.10$ & $1.33 \pm 0.18$ &  \\ \midrule

Di Scala \emph{et al.} \cite{scala2012} (simulation) & $2$ & $1.04 \pm 0.21$ & $0.29 \pm 0.03$ & $3.57 \pm 0.64$ & $1.04 \pm 0.04$  \\ \midrule

\begin{tabular}[c]{@{}l@{}} Bag \emph{et al.} \cite{bag2018}\\ (experiment; averaged results)  \hspace{1em} \end{tabular} & $3$  & $1.01 \pm 0.06$  & $0.41 \pm 0.02$  & $2.47 \pm 0.08$  & \\ \midrule

This study: non-interacting & $3$ & $1.35 \pm 0.11$ & $0.33 \pm 0.03$ & $4.1 \pm 0.10$ & $0.74 \pm 0.13$ \\

\hfil \quad interacting & $3$ & $2.41 \pm 0.24$ & $0.43 \pm 0.04$ & $5.6 \pm 0.20$ & $0.98 \pm 0.15$ \\ \midrule
\end{tabular}
\caption{Stationary critical scaling exponents for vortex depinning observed in several available numerical simulation and experimental studies.}
\label{table:steady_exp}
\end{table*}
With these estimated critical depinning forces $F_c$, we calculate the reduced drives $f$ and check if $v$ and $r$ scale respectively as per Eqs.~(\ref{eq:v_fin_tem_ans}) and  (\ref{eq:r_fin_tem_ans}). Employing global optimization methods \cite{wales1997, powell2003}, we have estimated the ensuing numerical values for the stationary critical exponents $\beta$, $\delta$, and $\nu$ that provide optimal scaling of the temperature- and drive-dependent observables $v$ and $r_g$, thereby facilitating convincing data collapse onto single master curves as demonstrated in Fig.~\ref{fig:crit_steadys_scaled}. 
\begin{table*}[htb]
\centering
\begin{tabular}{@{}llllll@{}}

\phantom{} \ \hspace{3em} & $b$ \hspace{5em} & $\lambda_C/z$ \hspace{3em} & $z$ \hspace{5em} & $\lambda_{C\,\textrm{ind}}$ \hspace{3em} & $\zeta_{\textrm{ind}}$ \\ \midrule[1.2pt]

Non-interacting vortices \hspace{1em} & $0.56 \pm 0.03$ & $0.61 \pm 0.02$ & $1.39 \pm 0.16$ & $0.85 \pm 0.10$ & $0.65 \pm 0.24$  \\ \midrule

Interacting flux lines & $0.29 \pm 0.03$ & $0.43 \pm 0.03$ & $1.43 \pm 0.15$ & $0.61 \pm 0.08$ & $0.98 \pm 0.16$  \\ \midrule

\end{tabular}
\caption{Critical aging and dynamical scaling exponents describing the non-equilibrium relaxation of vortices following critical drive quenches.}
\label{table:relax_exp}
\end{table*}

For the interacting vortex system, the optimal values of the exponents are found to be $\beta = 0.43 \pm 0.04$, $\delta = 5.6 \pm 0.2$, and $\nu = 0.98 \pm 0.15$ (Fig.~\ref{fig:crit_steadys_scaled}b/d). Our $\beta$ value shows good agreement with experiment  (Table~\ref{table:steady_exp}); however, our estimate for $\delta$ markedly differs from the value measured experimentally in Ref.~[\onlinecite{bag2018}]. In order to ascertain that our numerical data properly pertain to the asymptotic critical scaling regime, we have extracted the value of the product $\beta \delta$ using two distinct, complementary methods: (i) by scaling the $v$--$f$ curves for different temperatures giving $\beta \delta = 2.41 \pm 0.24$, and (ii) by scaling the $r_g$--$f$ curves yielding $\beta \delta = 2.5 \pm 0.2$; the two independent estimates show excellent agreement within our statistical and systematic error bars.

Likewise, we have evaluated the critical scaling exponents that yield excellent finite-temperature scaling for the non-interacting system to be  $\beta = 0.33 \pm 0.03$, $\delta = 4.1 \pm 0.1$, and $\nu = 0.74 \pm 0.13$ (Fig.~\ref{fig:crit_steadys_scaled}a/c). The values of $\beta \delta$ estimated, respectively, from the $v$--$f$ and $r_g$--$f$ scaling are $\beta \delta = 1.35 \pm 0.11$ and  $\beta \delta = 1.4 \pm 0.1$, which also agree nicely within our numerical errors.

Consequently, in both the non-interacting and interacting flux line systems, the fact that our estimates of $\beta \delta$ for scaling the $r_g$--$f$ curves using the ansatz (\ref{eq:r_fin_tem_ans}) are in agreement with the values obtained by scaling the $v$--$f$ curves with the extensively verified Eq.~(\ref{eq:v_fin_tem_ans}), in conjunction with the quality of data collapse for both scaling procedures, gives us confidence that we are properly accessing the asymptotic critical scaling regimes in either system. The discrepancies of our critical exponent values with those obtained in Ref.~\onlinecite{luo2007} might be caused by the lower number of $20$ layers along the magnetic field direction used in that study compared with our $L = 100$; perhaps for that smaller simulation domain thickness, the ultimate crossover to the two-dimensional scaling limit masks the asymptotic exponent values.

\section{Critical dynamics and aging scaling}

\begin{figure}
  \centering
  \includegraphics[width = .99\linewidth]{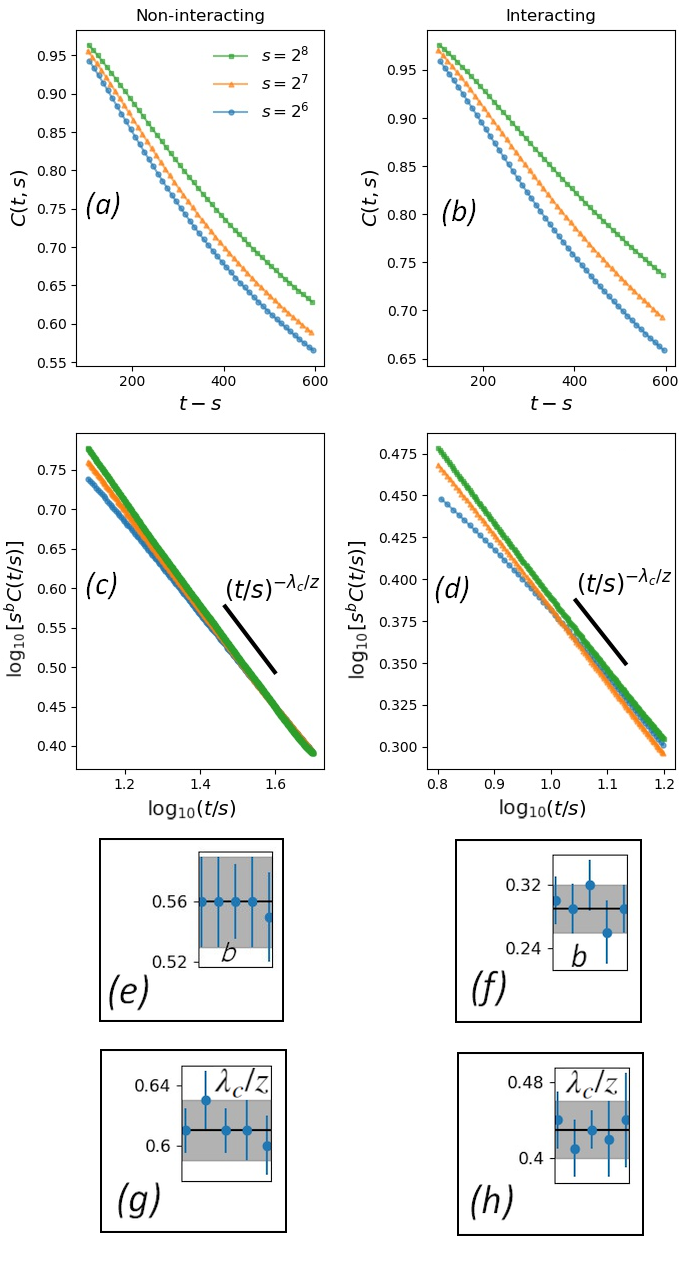}
	\caption{%
	(a, b) Two-time flux line height auto-correlation functions $C(t, s)$ following critical quenches at $T = 0.0005 \,\epsilon_0 b_0 / k_B$ for waiting times $s = 2^6t_0$, $2^7t_0$, and $2^8t_0$ as a functions of $t - s$; (c, d)  these auto-correlations scaled with $s^b$ as functions $t/s$ with $b = 0.56 \pm 0.03$ for (c) and $0.30 \pm 0.03$ for (d). Panels (a, c, e, g) and (b, d, f, h) represent the non-interacting and repulsively interacting flux line systems, respectively. The solid black line in (c, d) shows the power law dependence of the scaled quantities for $T = 0.0005 \epsilon_0 b_0 / k_B$ on $t/s$ with $\lambda_C / z = 0.61 \pm 0.015$ for (c) and $0.44 \pm 0.03$ for (d). Panels (e, f) and (g, h) respectively show the exponents $b$ and $\lambda_C/z$ estimated for critical quenches at five different temperatures $T= 0.0005$, $0.0006$, $0.0007$, $0.0008$, and $0.0009$ (left to right, in units of $\epsilon_0 b_0 / k_B$); the solid horizontal line in each panel represents the mean value of the data points and the shaded region indicates the error of the mean. The final mean exponent values are stated in Table~\ref{table:relax_exp}.
	\label{fig:crit_HAs}}
\end{figure}
In addition to finite-temperature critical scaling of one-time quantities near the depinning transition, we have studied the non-equilibrium relaxation of our flux line model systems following a drive quench from the moving state to the critical depinning regime. Investigating the two-time vortex height or transverse displacement auto-correlation function allows us to determine the associated dynamical and aging scaling exponents.

We begin by identifying the drive strength $F_m$ corresponding to the maximum steady-state radius of gyration for each temperature $T$ (Fig.~\ref{fig:crit_steadys} c/d). As explored in the preceding section, the gyration radius represents a good proxy for correlation length in the system, and it is reasonable to expect that its peak value must lie within the depinning drive regime. For critical quenches, we initially prepare the system in a moving  non-equilibrium steady state at high drive $F_d = 0.035\epsilon_0$ at the desired temperature $T$. Subsequently we suddenly switch to the depinning crossover drive $F_m(T)$, and start measuring two-time height auto-correlations $C(t, s)$ as the system relaxes from the quench over time. We perform these critical quench measurements for five different temperatures: $T = 0.0005$, $0.0006$, $0.0007$, $0.0008$, and $0.0009$ (values listed in units of $\epsilon_0 b_0 / k_B$).
\begin{figure}
  \centering
  \includegraphics[width = .99\linewidth]{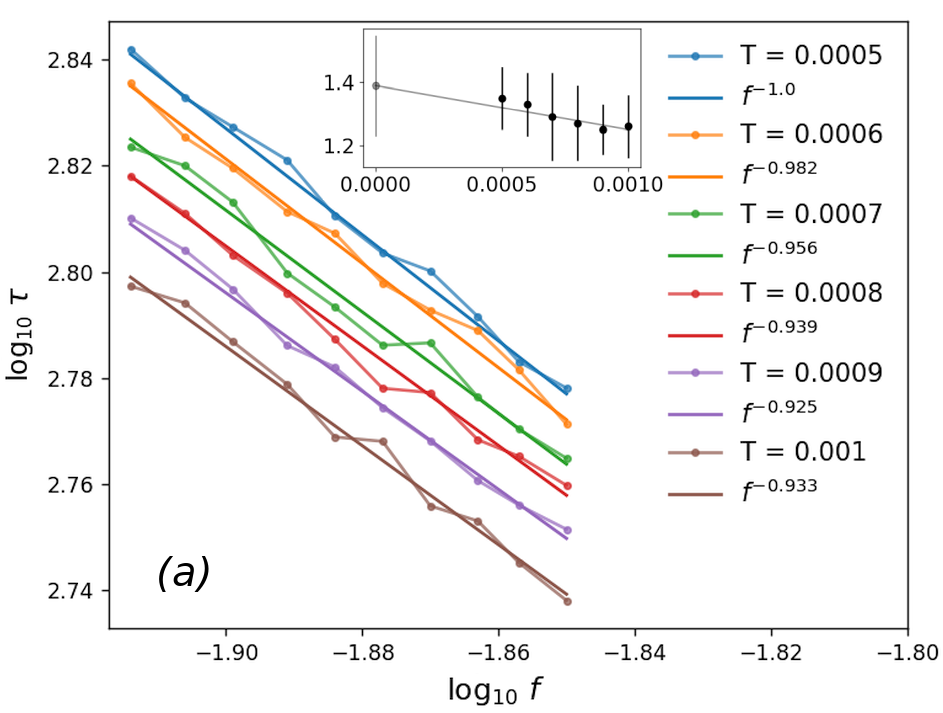}
  \includegraphics[width = .99\linewidth]{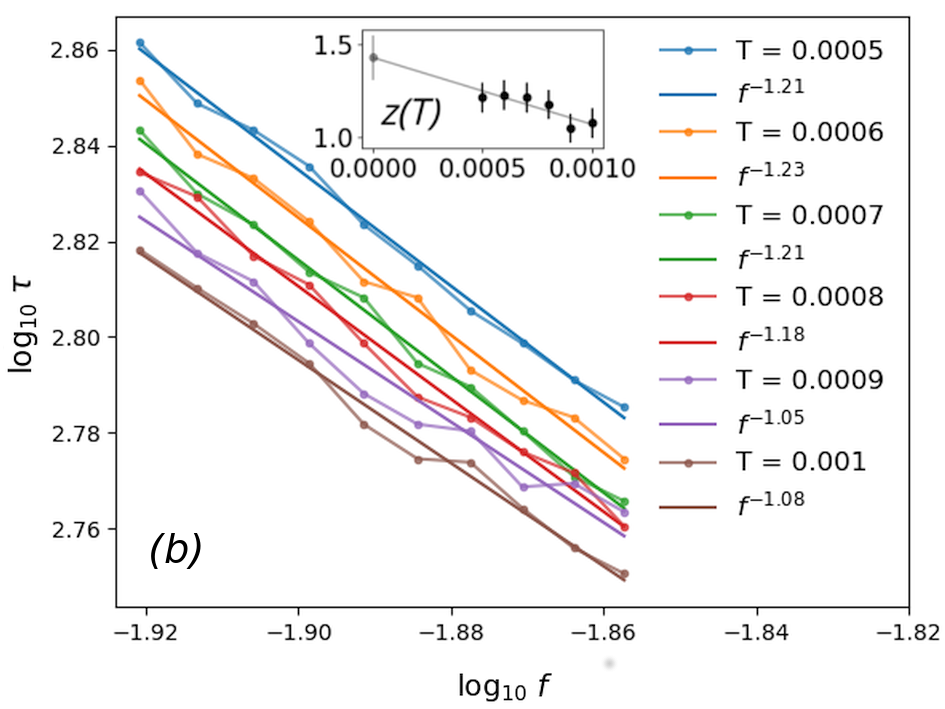}
	\caption{%
	Correlation time $\tau$ of (a) non-interacting and (b) interacting vortices as a function of drive $f$ (double-logarithmic scale) for six temperatures $T$ (values listed in units of $\epsilon_0 b_0 / k_B$) along with linear (power law) fits. The negative line slope for a given $T$ yields $\nu z$. In each panel, the inset shows the dynamic exponent $z$ (obtained from the $\log \tau$--$\log f$ data) as a function of $T$, along with a linear extrapolation to zero temperature. (Each data point originates from $1000$ independent simulation runs.)
	\label{fig:zs}}
\end{figure}
When we plot the height auto-correlations $C(t, s)$ against the time difference $t - s$ for different waiting times ($s = 2^6 t_0$, $2^7 t_0$, and $2^8 t_0$), we see clear breaking of time translation invariance (Fig.~\ref{fig:crit_HAs}a/b), the first indication of physical aging. The data are found to dynamically scale (Fig.~\ref{fig:crit_HAs}c/d) according to the full-aging ansatz (\ref{eq:aging_ansatz}). The scaled auto-correlations collapse on a master curve that appears to be linear on a double-logarithmic scale when plotted against $t/s$. This implies that the scaling function $f_C$ varies algebraically with $t/s$, indicating that the flux lines undergo simple aging after a critical quench. For long times, the master curve ultimately decays as a power law $(t/s)^{-\lambda_C/ z }$ where $\lambda_C$, and $z$ are respectively the auto-correlation and dynamic exponents. 

We have obtained excellent dynamical scaling collapse following critical drive quenches for our flux line system with and without vortex interactions for all five temperatures considered. Representative results for $T = 0.0005 \epsilon_0 b_0 / k_B$ are shown in Fig.~\ref{fig:crit_HAs}. Both in the absence or presence of repulsive interactions, the values of $\lambda_C / z$ and $b$ were found to agree (within statistical and systematic error bars) across all temperatures as seen in panels (e), (f), (g), and (h) of Fig.~\ref{fig:crit_HAs}. Indeed, in the critical scaling regime, at temperatures sufficiently close to zero and for $s \ll t$, one expects the aging scaling exponents to be universal  \cite{tauber2017, tauber2014, baumann2007, baumann2005, daquila2012, calabrese2006}. The observed universality of the aging scaling exponents for the temperatures considered here thus further supports the hypothesis of vortex depinning being a critical phenomenon (at zero temperature). Our extracted exponent values, averaged over the five different temperatures, are stated in Table~\ref{table:relax_exp}. Correlations for interacting vortices decay slower ($\lambda_C / z = 0.43 \pm 0.03$) than they do for non-interacting, independent flux lines ($\lambda_C / z = 0.61 \pm 0.02$) indicating that repulsive vortex-vortex interactions facilitate the formation of correlated vortex regions, and slow down the temporal relaxation of these collective deformations.

In order to obtain the dynamic critical exponent $z$, we first attempted to use a finite-temperature scaling ansatz as in our measurements of the static exponents $\beta$ and $\nu$. This approach failed, however, most likely on account of our not being able to get sufficiently close to the critical drive during the quenches. We then used an alternative method to evaluate $z$ which was to compute finite-temperature values of $z$ for multiple temperatures (Fig.~\ref{fig:zs}) in the following manner: For a given temperature $T$, we quenched moving systems to several drives $f > 0$ near $f = 0$ and computed the corresponding correlation times $\tau$; i.e,. the half life of $C(t, s = 128t_0)$. Since $\tau \sim f^{-\nu z}$, and with $\nu$ previously determined, we could thus infer $z$. We pursued this computation for six temperatures $T = 0.0005$, $0.0006$, $0.0007$, $0.0008$, $0.0009$, and $0.001$ (in units of $\epsilon_0 b_0 / k_B$). We then performed a linear extrapolation (Fig.~\ref{fig:zs} insets) to estimate the zero-temperature dynamic exponent, yielding $z = 1.39 \pm 0.16$ for non-interacting flux lines, and $z = 1.43 \pm 0.15$ for interacting vortices, indicating that the mutual repulsions induce slower critical relaxation. From the ratio $\lambda_C / z$ measured before, we may finally compute the auto-correlation exponents $\lambda_C = 0.85 \pm 0.10$ for the non-interacting system, while $\lambda_C = 0.61 \pm 0.08$ for interacting flux lines. All our results for the dynamical and aging scaling exponents are summarized in Table~\ref{table:relax_exp}.

Hyperscaling relations connecting the growth exponent $\nu$, the order parameter exponent $\beta$, the roughness exponent $\zeta$, and the dynamic critical exponent $z$ have been derived, the latter using statistical tilt symmetry \cite{nattermann1992, narayan1992, le_doussal2002}:
\begin{equation} \label{eq:hypsc_rel}
\nu = 1/(2 - \zeta) \ \textrm{, } \ \beta = (z - \zeta) \nu \ .
\end{equation}
Using our numerically obtained values of $\nu$ and $\beta$ (Table~\ref{table:steady_exp}), we can compute $\zeta = 2 - 1 / \nu$ and $z = \zeta + \beta / \nu$. We find $\zeta = 0.65 \pm 0.24$, $z = 1.10 \pm 0.36$ for non-interacting vortices, whereas $\zeta = 0.98 \pm 0.16$, $z = 1.42 \pm 0.25$ with repulsive interactions present. For the interacting vortices, the value for the dynamical exponent from the hyperscaling relations (\ref{eq:hypsc_rel}) is thus fully consistent with our direct numerical estimate listed in Table~\ref{table:relax_exp}; for the non-interacting lines, we observe a larger deviation, but still well within our error bars.

\section{Conclusions}
In this detailed numerical study, we have employed a coarse-grained three-dimensional elastic line model of magnetic vortices and overdamped Langevin molecular dynamics simulations to investigate the critical depinning of flux lines from randomly distributed weak attractive point pinning centers. We have performed finite-temperature scaling of one-time quantities, namely the mean vortex velocity and flux line gyration radius, to obtain consistent estimates of the stationary critical exponents $\beta$, $\delta$, and $\nu$. Independent analyses of data collapse for these observables confirm that we are properly accessing the asymptotic critical scaling regime in both systems of non-interacting flux lines and mutually repulsive vortices. Our estimate for the correlation length exponents $\nu$ in three dimensions turns out remarkably close to, but slightly smaller than the numerical result from Ref.~\onlinecite{scala2012} obtained via finite-size scaling for a two-dimensional vortex system. Our value for $\beta$ is in very good agreement with recent experimental results for 2H-NbS$_2$ single crystals \cite{bag2018}. However, our estimate for $\delta$ clearly deviates from the corresponding measured value.

Furthermore, we have investigated dynamic scaling properties in the non-equilibrium relaxation regime following drive quenches from the moving vortex state into the critical depinning regime, and thus determined the aging scaling exponent $b$, auto-correlation exponent $\lambda_C$, and dynamic critical exponent $z$ for the relaxation of the system via the analysis of two-time flux line height auto-correlation functions and the aid of hyperscaling relations between the static and dynamic exponents. We found evidence for universal scaling near the depinning threshold in the form of temperature independence of the aging scaling exponents indicating that we are accessing the critical aging regime in the system, and providing further support for elastic depinning constituting a dynamic critical phenomenon. Mutual repulsive interactions collectively cage flux lines and hence slow down the decay of correlations in the system as evidenced by the smaller value of $\lambda_C / z$ compared to the relaxation of non-interacting vortices.
\bigskip

\section*{Acknowledgments} \label{acknowledgements}
This material is based upon work supported by the U.S. Department of Energy, Office of Science, Office of Basic Energy Sciences, Division of Materials Sciences and Engineering under Award Number DE-SC0002308.

\bibliographystyle{apsrev4-1}
\bibliography{bibliography}

\end{document}